\documentstyle[12pt]{article}

\def\thebibliography#1{\bigskip\section*{}\bigskip\list
{$^{\arabic{enumi}}$}{\settowidth\labelwidth{#1}\leftmargin\labelwidth
\advance\leftmargin\labelsep
\usecounter{enumi}}
\def\newblock{\hskip .11em plus .33em minus .07em}
\sloppy\clubpenalty4000\widowpenalty4000
\sfcode`\.=1000\relax}

\def\op#1{\mathop{\fam0 #1}\limits}

\newcommand{\beq}{\begin{equation}}
\newcommand{\eeq}{\end{equation}}
\newcommand{\ben}{\begin{eqnarray}}
\newcommand{\een}{\end{eqnarray}}
\newcommand{\be}{\begin{eqnarray*}}
\newcommand{\ee}{\end{eqnarray*}}
\newcommand{\bea}{\begin{eqalph}}
\newcommand{\eea}{\end{eqalph}}

\newcommand{\bC}{{\bf C}}

\newcommand{\cA}{{\cal A}}
\newcommand{\cG}{{\cal G}}

\newcommand{\cH}{{\cal H}}

\newcommand{\bq}{{\bf q}}
\newcommand{\bb}{{\bf 1}}
\newcommand{\bp}{{\bf p}}
\newcommand{\br}{{\bf r}}

\newcommand{\bR}{{\bf R}}
\newcommand{\al}{\alpha}
\newcommand{\bt}{\beta}
\newcommand{\dl}{\delta}
\newcommand{\la}{\lambda}

\newcommand{\f}{\phi}
\newcommand{\vf}{\varphi}
\newcommand{\om}{\omega}

\newcommand{\m}{\mu}

\newcommand{\g}{\gamma}

\newcommand{\ve}{\varepsilon}

\newcommand{\bom}{{\bf\Omega}}
\newcommand{\w}{\wedge}

\newcommand{\wh}{\widehat}
\newcommand{\ol}{\overline}
\newcommand{\dr}{\partial}

\newcommand{\ot}{\otimes}

\newcommand{\lng}{\langle}
\newcommand{\rng}{\rangle}

\let\ssection=\section
\renewcommand{\section}{\setcounter{equation}{0}\ssection}

\newcounter{eqalph}
\newcounter{equationa}
\newcounter{theorem}
\newcounter{proposition}
\newcounter{lemma}
\newcounter{corollary}
\newcounter{definition}

\newenvironment{eqalph}{\stepcounter{equation}
\setcounter{equationa}{\value{equation}}
\setcounter{equation}{0}

\begin{eqnarray}}{\end{eqnarray}\setcounter{equation}{\value{equationa}}}

\def\thedefinition{\arabic{definition}}

\newenvironment{rem}{\medskip\noindent{\it Remark:}}{\medskip}

\newcommand{\mar}[1]{}

\begin{document}
\hbox{}

{\parindent=0pt

{\large\bf Quantum Jacobi fields in Hamiltonian  mechanics}
\bigskip

{\sc G.Giachetta}\footnote{Electronic mail:
giachetta@campus.unicam.it} {\sc and L.Mangiarotti}\footnote{Electronic mail:
mangiaro@camserv.unicam.it}

{\sl Department of Mathematics and Physics, University of Camerino, 62032
Camerino (MC), Italy}
\medskip

{\sc G. Sardanashvily}\footnote{Electronic mail:
sard@grav.phys.msu.su}

{\sl Department of Theoretical Physics, 
Moscow State University, 117234 Moscow, Russia}
\bigskip
 
Jacobi fields of classical solutions of a Hamiltonian
mechanical system are quantized in the framework of the vertical-extended
Hamiltonian formalism. Quantum Jacobi fields characterize
quantum transitions between classical solutions.}
\bigskip


\noindent 
{\bf I. INTRODUCTION}
\bigskip

One of the main problems in 
algebraic quantum theory is to describe transitions between non-equivalent
states of an algebra $A$ of observables of a quantum system. If $A$ is a
$C^*$-algebra, one can consider the enveloping von Neumann algebra $B$ of $A$
and find an (adjoint) element $T$ of its center
such that non-equivalent states of $A$ appear to be (generalized)
eigenstates of $T$ with different eigenvalues.$^1$ The $T$ exemplifies a
superselection operator.$^2$ Superselection operators are usually associated
to macroscopic (classical) observables because they commute with all elements
of an algebra of a quantum system. Furthermore, one may hope that there exists
an extended quantum system whose algebra includes both superselection operators
and operators which  transform  non-equivalents states of the algebra
$A$. For instance, let us mention quantization over
different classical background fields in quantum field theory.
Here, we are concerned with the similar problem in (non-autonomous) 
Hamiltonian mechanics.

Given a classical Hamiltonian mechanical system, one can associate to its
solutions  and their Jacobi fields the Hermitian operators
$\br$ and $\dot \br$  in a Hilbert space such that the eigenvalues of the
mutually commutative operators of classical solutions
$\br$ are values $r(t)$  of these solutions at 
instant $t$. The Jacobi field operators $\dot \br$ perform a transition
between the eigenstates of operators of solutions
$\br$. Operators of solutions play the role of superselection operators 
if one considers the standard quantization of linear deviations of
the above Hamiltonian system over a classical solution.  The Hamiltonian
of these deviations depends on the operators of solutions $\br$ seen as
$c$-numbers with respect to the deviation operators, but the Jacobi field
operators $\dot\br$ acting on this Hamiltonian perform the transition between
quantizations over different classical solutions. 

The key point of the above quantization scheme is the particular commutation
relations of operators of solutions and Jacobi field operators. They result
from the Poisson bracket of a classical Hamiltonian system which is extended in
order to include Jacobi fields as follows.

A generic momentum phase space of a  Hamiltonian
mechanical system is a Poisson fiber bundle $\Pi\to \bR$ over the time axis
$\bR$.$^{3,4}$ We restrict our consideration to mechanical systems which
admit a configuration space (see Ref. [5] for opposite examples). This is 
a smooth fiber bundle
$Q\to \bR$ which is
provided with bundle coordinates  
$(t,q^k)$ where $t$ is a fixed Cartesian coordinate on $\bR$.
Of course, the
fiber bundle $Q\to\bR$ is trivial, but its different trivializations $Q\cong
\bR\times M$ correspond to different reference frames. Therefore, we deal
with local fiber coordinates $(q^k)$ subject to time-dependent transformations
in order to study to what extent the quantization procedure
below is frame-independent. 
The momentum phase space of Hamiltonian mechanics on the configuration bundle
$Q\to\bR$ is the vertical cotangent bundle
$\Pi=V^*Q$ of
$Q\to\bR$, endowed with the holonomic coordinates $(t,q^k,p_k=\dot q_k)$.
This momentum phase space admits the canonical exterior three-form 
\be
\bom=dp_k\w dq^k\w dt, 
\ee
which provides $V^*Q$ with the canonical Poisson structure$^{4,6}$
\be
\{f,g\}=\dr^kf\dr_kg-\dr_kf\dr^kg, \qquad f,g\in C^\infty(V^*Q). 
\ee
A Hamiltonian system on $V^*Q$ is
defined by a Hamiltonian form
\mar{b4210}\beq
H= p_k dq^k -\cH (t,q^k,p_k) dt, \label{b4210}
\eeq
which leads to 
the Hamilton equations
\mar{m41}\beq
d_t q^k =\dr^k\cH, \qquad  d_t p_k =-\dr_k\cH. \label{m41}
\eeq

Note that any Poisson bundle $\Pi\to\bR$ (i.e., the fibration
$\Pi\to\bR$ is a symplectic foliation of
$\Pi$) is locally isomorphic to the above case $\Pi=V^*Q$.$^{4,7}$ 

There are different approaches to mathematical definition of Jacobi fields
in Lagrangian and Hamiltonian dynamics.$^{5,8-11}$

To describe Jacobi fields of solutions of the Hamilton equations
(\ref{m41}), we consider the extension of a Hamiltonian system on
$Q$ to the vertical tangent bundle $VQ$ of $Q\to\bR$, 
viewed as a new vertical-extended configuration space.$^{4,12}$
It is provided with the
holonomic coordinates $(t,q^k,\dot q^k)$. 
The corresponding momentum phase space is the vertical cotangent bundle 
$V^*VQ$ of  
$VQ\to \bR$. It is canonically isomorphic to the
vertical tangent bundle 
$VV^*Q$  of the momentum phase space
$V^*Q\to\bR$, and is coordinated by  $(t, q^k, p_k,\dot q^k,\dot p_k)$.
One obtains easily from the coordinate
transformation laws that $(q^k, \dot p_k)$ and $(\dot q^k, p_k)$ are
canonically conjugate pairs.

The above mentioned isomorphism $V^*VQ\cong VV^*Q$  enables
one to extend a Hamiltonian system on $Q$ to $VQ$ as its vertical prolongation
by means of the the vertical tangent functor 
\be
\dr_V=\dot q^k\dr_k +\dot p_k\dr^k.
\ee
Namely, the vertical momentum phase space $VV^*Q$ admits the 
canonical three-form
\be
\bom_V=\dr_V\bom=[d\dot p_k\w dq^k +dp_k\w d\dot q^k]\w dt. 
\ee
It provides $VV^*Q$ with the canonical Poisson structure
\mar{j4}\beq
\{f,g\}_V=\dot\dr^kf\dr_kg +\dr^kf\dot\dr_kg-\dot\dr_kf\dr^kg-\dr_kf\dot\dr^kg,
\qquad f,g\in C^\infty(VV^*Q),
\label{j4}
\eeq
where the compact notation $\dot\dr_k=\dr/\dr \dot q^k$,
$\dot\dr^k=\dr/\dr \dot p_k$ is used.
The vertical extension of the Hamiltonian form $H$ (\ref{b4210}) reads
\mar{j5}\beq
H_V=\dot p_kdq^k + p_k d\dot q^k-\cH_Vdt, \qquad \cH_V=\dr_V\cH=
(\dot q^k\dr_k +\dot p_k\dr^k)\cH. \label{j5}
\eeq
It leads to the Hamilton equations 
\mar{z740}\bea
&& d_t q^k=\dot\dr^k\cH_V =\dr^k\cH, \qquad 
 d_t p_k=-\dot\dr_k\cH_V =-\dr_k\cH,\label{z740a}\\
&& d_t \dot q^k=\dr^k\cH_V=\dr_V\dr^k\cH,\qquad
d_t\dot p_k=-\dr_k\cH_V=-\dr_V\dr_k\cH,\label{z740b}
\eea
where the equations (\ref{z740a}) are exactly the Hamilton
equations (\ref{m41}) of the
original Hamiltonian system. Given a solution $r(t)$ of the
Hamilton equations (\ref{z740a}), let 
$\dot r(t)$ be a Jacobi field, i.e., 
$r(t)+\ve \dot r(t)$, $\ve\in\bR$,
is also a solution of the same Hamilton equations modulo terms of order more
than one in
$\ve$. Then the Jacobi field $\dot r(t)$ fulfills 
the Hamilton equations  (\ref{z740b}).

In particular, let $r$ be a local solution
of the Hamilton equations (\ref{z740a}), given by local functions 
\mar{j40}\beq
r^k(t,a^j, b_j), \qquad  r_k(t,a^j, b_j), \qquad a^j=r^j(0), \qquad
b_j=r_j(0). \label{j40}
\eeq
 Its Jacobi field $\dot r$ is a solution of
the Hamilton equations(\ref{z740b}), represented by local functions
\mar{j41}\beq
\dot r^k(t,a^j,b_j,c^j,s_j), \qquad \dot r_k(t,a^j,b_j,c^j,s_j), \qquad  
c^j=\dot r^j(0), \qquad s_j=\dot r_j(0), \label{j41}
\eeq
which fulfill the system of linear
ordinary differential equations
\mar{j42}\beq
\dr_t\dot r^k=\dot r^j(\dr_j\dr^k\cH)(t,r) + \dot
r_j(\dr^j\dr^k\cH)(t,r), \quad
\dr_t\dot r_k=-\dot r^j(\dr_j\dr_k\cH)(t,r) - \dot
r_j(\dr^j\dr_k\cH)(t,r). \label{j42}
\eeq
These equations can be written in the matrix form 
\be
\dr_t(\dot r^k,\dot r_k)= (\dot r^k,\dot r_k) M(t, a^j, b_j).
\ee
Then the Jacobi field $\dot r$ (\ref{j41}) can be written as the
time-ordered matrix exponent
\mar{j24}\beq
(\dot r^k,
\dot r_k)=(c^k,s_k) T \exp\left[\op\int_0^t  M(t', a^j, b_j) dt'\right].
\label{j24}
\eeq

\begin{rem}
The similar vertical extension of Lagrangian mechanics on the jet
manifold $J^1Q$ of $Q$ to the jet manifold $J^1VQ\cong VJ^1Q$ of $VQ$
provides a description of Jacobi fields of solutions of the Euler--Lagrange
equations.$^4$  Given a Lagrangian $L$ on the velocity phase space $J^1Q$
coordinated by
$(t,q^k,q^k_t)$, its extension to
$VJ^1Q$ reads
\be
L_V=(\dot q^k\dr_k +\dot q^k_t\dr^t_k)L.
\ee
The same procedure
is appropriate for Lagrangian and Hamiltonian field theory.$^{13}$
It is also a preliminary step toward the SUSY extension of field
theory and time-dependent mechanics.$^{4,13-15}$
\end{rem}

 We aim to quantize the vertical-extended Hamiltonian system when
$Q\to\bR$ is a vector bundle and a Hamiltonian $\cH$ is a polynomial of
coordinates and momenta.  
Since $Q\to\bR$ is a vector bundle and $VV^*Q\to \bR$ is so, it is a
particular variant of geometric quantization which reduces to the canonical
quantization of the Poisson bracket (\ref{j4}). 
The physical peculiarity of quantization of a vertical-extended
Hamiltonian system lies in the form of the Poisson bracket (\ref{j4}). To
illustrate this peculiarity, let us follow to naive canonical quantization and
assign to quantities $q^k$, $p_k$,
$\dot q^k$, $\dot p_k$  the operators
$\bq^k$,
$\bp_k$, $\dot \bq^k$, $\dot \bp_k$ which 
satisfy the canonical commutation relations
\mar{j9}\bea
&& [\bq^k,\bq^j]=[\bp_k,\bp_j]=[\bp_k,\bq^j]=0, \label{j9a}\\
&& [\dot\bq^k,\dot\bq^j]=[\dot\bp_k,\dot\bp_j]=[\dot\bp_k,\dot\bq^j]=0,
\label{j9b}\\
&& [\bq^k,\dot\bq^j]=[\bp_k,\dot\bp_j]=0, \qquad [\bp_k,\dot\bq^j]=
[\dot\bp_k,\bq^j] =-i\hbar \dl^j_k\bb \label{j9c}
\eea
as the operator form of the Poisson bracket (\ref{j4}). The
commutation relations (\ref{j9a}) show that operators $\bq^k$, $\bp_k$ of
the original Hamiltonian system mutually commute and,
consequently, characterize compatible observables. 

\begin{rem}
Recall the well-known method of second quantization in field theory,
which is applied both to quantization of free fields$^{16}$ and, especially,
to quantization in the presence of a back-ground field.$^{17-19}$ Given a
classical field equation, one considers a complete
set of its solutions and associates to each solution at any instant
the  creation and annihilation operators or canonically conjugate
operators which obey the canonical commutation relations. In the case under
consideration, the Poisson bracket (\ref{j4}) makes the operators of
classical solutions commutative.
\end{rem}

Furthermore,
the commutation relations  (\ref{j9b}) show that the Jacobi field operators
$\dot\bp_k$,
$\dot\bq^k$ are not canonically conjugate, in contrary to the customary
quantization of linear deviations of a classical system. Indeed, if
$Q\to\bR$ is a vector bundle, there is the canonical bundle
isomorphism
\mar{j15}\beq
VV^*Q\cong Q\times Q^*\times Q\times Q^* \label{j15}
\eeq
over $\bR$. Therefore, the vertical momentum phase space $VV^*Q$ can admit
both the Poisson structure (\ref{j4}) and the Poisson structure
\be
\{f,g\}'_V=\dr^kf\dr_kg - \dr_kf\dr^kg
+\dot\dr^kf\dot\dr_kg-\dot\dr_kf\dot\dr^kg.
\ee
This Poisson structure leads to the canonical commutation relations
for quantum linear deviations. 

To provide a unified scheme of quantization both of Jacobi fields and linear
deviations of a classical Hamiltonian system, let us consider its second
vertical extension to the second vertical tangent bundle $V^2Q$. The $V^2Q$
is a subbundle of the repeated vertical tangent bundle $VVQ$, given by the
coordinate relations $\dot q^k= Q^k$ where $(t,q^k,\dot q^k, \dot{\rm
q}^k,\ddot q^k)$ are holonomic coordinates on $VVQ$. Due to the canonical
isomorphism
$V^*V^2Q\cong V^2V^*Q$, the second vertical momentum phase space $V^*V^2Q$
is equipped with the holonomic coordinates 
$(t,q^k,p_k, \dot q^k, \dot p_k, \ddot q^k,\ddot p_k)$, and
one can extend a Hamiltonian system from $VQ$ to $V^2Q$ by means of the second
vertical tangent functor 
\be
\dr_{V^2}=\dot q^k\dr_k + \dot p_k\dr^k +\ddot q^k\dot \dr_k +\ddot p_k\dr^k.
\ee
Namely, the second vertical momentum phase space $V^2V^*Q$ admits the
canonical three form
\be
\bom_{V^2}=\dr_{V^2}\bom_V=[d\ddot p_k\w dq^k + dp_k\w\ddot q^k + 2d\dot p_k
\w\dot q^k]\w dt.
\ee
It provides $V^2V^*Q$ with the Poisson bracket 
\be
\{f,g\}_{V^2}=\ddot\dr^kf\dr_kg
+\dr^kf\ddot\dr_kg-\ddot\dr_kf\dr^kg-\dr_kf\ddot\dr^kg
+\frac12(\dot\dr^kf\dot\dr_kg - \dot\dr_kf\dot\dr^kg),
\ee
where $(q^k,\ddot p_k)$, $(\ddot q^k,p_k)$, $(Q^k=\sqrt2\dot
q^k,P_k=\sqrt2\dot p_k)$ are canonically conjugate pairs. 
This Poisson bracket leads to the following commutation relations of naive
canonical quantization 
\be
&& [\bq^k,\bq^j]=[\bp_k,\bp_j]=[\bp_k,\bq^j]=0,\\
&& [\ddot\bq^k,\ddot\bq^j]=[\ddot\bp_k,\ddot\bp_j]=[\ddot\bp_k,\ddot\bq^j]=0,
\\
&& [\bq^k,\ddot\bq^j]=[\bp_k,\ddot\bp_j]=0, \qquad [\bp_k,\ddot\bq^j]=
[\ddot\bp_k,\bq^j] =-i\hbar \dl^j_k\bb,\\
&& [Q^k,Q^j]=[P_k,P_j]=0, \qquad [P_k,Q^j] =-i\hbar \dl^j_k\bb. 
\ee
Comparing these commutation relations with the commutation relations
(\ref{j9a}) -- (\ref{j9b}), we can treat $\ddot\bq$, $\ddot\bp$ as quantum
Jacobi fields, while $P$, $Q$ obey the canonical commutation relations
 for quantum deviations.
Their common Hamiltonian reads
\mar{j5'}\ben
&& \cH_V=\dr_{V^2}\dr_V\cH=(\ddot q^k\dr_k +\ddot p_k\dr^k)\cH +
\frac12(Q^k\dr_k +P_k\dr^k)^2\cH =\nonumber \\
&&\cH_1(t,q^k,p_k,\ddot q^k,\ddot
p_k) +
\cH_2(t,q^k,p_k,P^k,P_k). \label{j5'}
\een
If the original Hamiltonian $\cH$ is quadratic in coordinates
$q$ and momenta $p$, Jacobi fields with the Hamiltonian $\cH_1$ and 
linear deviations with the Hamiltonian $\cH_2$ are quantized independently.
In a general case, since linear deviations are treated as pure quantum
objects, the system of classical solutions and their Jacobi fields with the
Hamiltonian $\cH_1$ is quantized at first. Then one quantizes linear 
deviations whose Hamiltonian contains superselection operators of classical
solutions.  

This quantization scheme can also be applied to canonical
quantization of field theory seen as an instantaneous Hamiltonian system (in
the spirit of Ref. [8,20]).

\bigskip 

\noindent 
{\bf II. THE PREQUANTIZATION ALGEBRA}
\bigskip

In this Section, we construct an involutive algebra, called the
prequantization algebra, whose Hermitian elements can represent classical
solutions of the original Hamiltonian system and quantum Jacobi fields.

Let $C^\infty(VV^*Q)$ be a ring of smooth real functions on the
vertical momentum phase space 
$VV^*Q$. It is a real Lie algebra  with
respect to the Poisson bracket (\ref{j4}), called the Poisson algebra.
Quantization on a Poisson manifold usually implies an assignment of a
Hermitian operator
$\wh f$ to each element $f\in C^\infty(VV^*Q)$ such that the Dirac condition
\mar{qm514}\beq
[\wh f,\wh f']=-i\hbar\wh{\{f,f'\}_V} \label{qm514}
\eeq
holds. One can follow the geometric quantization procedure. 
Then, since $Q\to \bR$ is a vector bundle and since the Poisson
bracket (\ref{j4}) does not contain a derivation with respect to time, 
we can
easily observe that the geometric quantization of the Poisson algebra 
$C^\infty(VV^*Q)$ reduces
to the canonical quantization of its Poisson subalgebra $\cA$ of functions
which are affine in fiber coordinates $q^k$, $p_k$,
$\dot q^k$, $\dot p_k$ on $VV^*Q\to\bR$.$^{21}$ Therefore, we will simply
provide this canonical quantization in a straightforward manner in Section IV.

Since the Poisson algebra $\cA$ can be quantized, let us use its isomorphism
as a $C^\infty(\bR)$-module to the module 
$E(\bR)$ of global section of the Whitney sum
\be
E=VV^*Q^*\oplus \bR^2
\ee
of the line bundle
$\bR^2\to\bR$ and the dual $VV^*Q^*$ of $VV^*Q$.
This isomorphism is given by the assignment  
\mar{j12}\ben
&& \cA\ni f=  a_k(t)q^k + b^k(t)p_k + c_k(t) \dot
q^k + d^k(t)\dot p_k + e(t) \mapsto\label{j12} \\
&& \qquad  a_k(t)\vf^k + b^k(t) \chi_k + c_k(t) \dot \vf^k +
d^k(t)\dot \chi_k +e(t)I\in E(\bR), \nonumber
\een
where $I$ is the fiber basis for 
$\bR^2\to\bR$, while
$\{\vf^k\}$ are fiber bases for the dual vector bundle $Q^*\to\bR$
coordinated by $(q_k)$, and 
\be
\{\chi_k=\ol dq_k,\quad \dot
\vf^k=\dr/\dr q_k,\quad \dot
\chi_k=\ol d\dot q_k\}
\ee
are holonomic fiber bases for the vector bundle
$VV^*Q^*\to Q^*$.
Due to the isomorphism (\ref{j12}), the module $E(\bR)$ inherits from
$\cA$ the structure of a Lie 
$C^\infty(\bR)$-algebra with local generators
$\{\vf^k,\chi_k,\dot\vf^k,\dot\chi_k, I\}$ subject to the fiberwise 
Lie algebra multiplication
\mar{j13}\bea
&& \{\vf^k, \vf^j\}_E=\{\chi_k,\chi_j\}_E=\{\chi_k,\vf^j\}_E=0, \label{j13a}\\
&&
\{\dot\vf^k,\dot\vf^j\}_E=\{\dot\chi_k,\dot\chi_j\}_E=\{\dot\chi_k,\dot\vf^j\}_E=0,
\label{j13b}\\
&& \{\vf^k,\dot\vf^j\}_E=\{\chi_k,\dot\chi_j\}_E=0, \qquad
\{\chi_k,\dot\vf^j\}_E=
\{\dot\chi_k,\vf^j\}_E =\dl^j_kI, \label{j13c}
\eea
where $I$ commutes with all elements.
This Lie algebra structure is coordinate-independent because
linear transformations of the fiber bases $\{\vf_k\}$ for
the vector bundle $Q^*\to\bR$ and the induced holonomic transformations of
the fiber bases $\{\chi_k,\dot\vf^k,\dot\chi_k\}$ for 
$VV^*Q^*\to Q^*$ maintain the multiplication relations (\ref{j13a}) --
(\ref{j13c}). Therefore, this fiberwise multiplication
makes $E\to \bR$ a fiber bundle of Lie algebras.
Its fiber $E_t$ over a point $t\in \bR$ is a real Lie
algebra given by the multiplication relations (\ref{j13a}) -- (\ref{j13c})
with respect to the holonomic frame $\{\vf^k,\chi_k,\dot\vf^k,\dot\chi_k, I\}$
at $t\in\bR$. The Lie algebra $C^\infty(\bR)$-algebra of global sections of
$E$, by construction, is isomorphic to the above mentioned  Poisson subalgebra
$\cA$. 

Let us consider the 
enveloping algebra $\ol E_t$ of the Lie algebra $E_t$ for each $t\in \bR$. 
It is the quotient of tensor algebra
\be
\bR \op\oplus_{m=1}(\op\ot^m E_t)
\ee
with respect to the two-sided ideal generated by
all elements of the form 
\be
e\ot e'-e'\ot e -\{e,e'\}_E, \qquad e,e'\in E_t.
\ee
There is the canonical monomorphism $E_t\to \ol E_t$ such that
the basis for $E_t$ is also a basis for $\ol E_t$.$^{22}$ Then, identifying
$E_t$ to its image in
$\ol E_t$, one can write the Lie algebra product in $E_t$ as the commutator
\be
\{e,e'\}_E=[e,e']=ee'-e'e
\ee 
with respect to the product in $\ol E_t$. 

 Let us complexify $\ol E_t$ and $E_t$ as $\ol\cG_t=\ol E_t\ot_\bR\bC$ and
 $\cG_t= E_t\ot_\bR\bC$, respectively.  Since the frames
$\{\vf^k\}$, $\{\chi_k\}$, $\{\dot\vf^k\}$, $\{\dot\chi_k\}$ are
transformed independently due to the splitting (\ref{j15}), one can choose a
basis
\mar{j17}\beq
\{\vf^k,\pi_k=-i\hbar\chi,\dot\vf^k,\dot\pi_k=-i\hbar\dot\chi_k, I\}
\label{j17}
\eeq
for $\cG_t$ such that
\mar{j16}\bea
&& [\vf^k, \vf^j]=[\pi_k,\pi_j]=[\pi_k,\vf^j]=0, \label{j16a}\\
&&
[\dot\vf^k,\dot\vf^j]=[\dot\pi_k,\dot\pi_j]=[\dot\pi_k,\dot\vf^j]=0,
\label{j16b}\\
&& [\vf^k,\dot\vf^j]=[\pi_k,\dot\pi_j]=0, \qquad
[\pi_k,\dot\vf^j]=
[\dot\pi_k,\vf^j] =-i\hbar\dl^j_kI. \label{j16c}
\eea
where $I$ commutes with all elements.
 Of course, the frames $\{\pi_k\}$ and $\{\dot\pi_k\}$ possess the same
holonomic transformations as $\{\chi_k\}$ and $\{\dot\chi_k\}$. 

The complex enveloping algebra $\ol\cG_t$ can be provided with an involution
$*$.$^{22}$  Given a basis $\{\vf^k,\pi_k,\dot\vf^k,\dot\pi_k,I\}$ (\ref{j17})
of
$\ol\cG_t$, this operations is
defined by the the following conditions: 
\begin{itemize} \begin{enumerate}
\item  the basis elements $\vf^k$, $\pi_k$, $\dot\vf^k$, $\dot\pi_k$, $I$ 
are invariant with respect to the involution $*$, 
\item $\la^*=\ol
\la$ for any complex number $\la$, 
\item $(e_1\cdots e_m)^*=e^*_m\cdots e^*_1$ for all elements $e_1,\ldots,
e_m$ of $\ol\cG$.
\end{enumerate} \end{itemize}
The condition (iii) is well-defined because it together
with the condition (i) imply that the involution $*$ maintains the commutation
relations (\ref{j16a}) -- (\ref{j16c}). It also follows that the
definition of $*$ is coordinate-independent. The involution $*$ makes
$\ol \cG_t$ an involutive algebra generated by the Hermitian basis 
$\{\vf^k,\pi_k,\dot\vf^k,\dot\pi_k,I\}$. Holonomic transformations of this
basis yield automorphisms of $\ol\cG_t$.

The involutive algebras $\ol\cG_t$, $t\in \bR$ make up the fiber
bundle
$\ol\cG\to
\bR$ of involutive  algebras over $\bR$. Global sections of
this fiber bundle $\ol\cG$ constitute an involutive algebra $\ol\cG(\bR)$ over
the ring
$\bC^\infty(\bR)$ of smooth complex functions on $\bR$. There is
$C^\infty(\bR)$-module monomorphism
\mar{j18}\ben
&& \cA\ni f=  a_k(t)q^k + b^k(t)p_k + c_k(t) \dot
q^k + d^k(t)\dot p_k + e(t) \mapsto\label{j18} \\
&& \qquad  a_k(t)\vf^k + b^k(t) \pi_k + c_k(t) \dot \vf^k +
d^k(t)\dot \pi_k +e(t)I=\wh f\in \ol \cG(\bR), \nonumber
\een
which assigns a Hermitian element $\wh f$ of the involutive algebra
$\ol\cG(\bR)$ to each element $f$ of the Poisson subalgebra $\cA$ in accordance
with the prequantization Dirac condition (\ref{qm514}). 

The monomorphism (\ref{j18}) can be extended as
\mar{j21}\beq
C^\infty(VV^*Q)\supset C^\infty(V^*Q) \ni f(t,q_j,p^j)\mapsto
f(t,\vf_j,\pi^j)=\wh f\in
\ol\cG(\bR)  
\label{j21}
\eeq
to functions on $VV^*Q$ which
are the pull-back of polynomial functions
$f$ of fiber coordinates 
$q^k$, $p_k$ on $V^*Q\to\bR$. 
Furthermore, we add to $\ol\cG(\bR)$ the images $\wh f$ (\ref{j21})
of functions $f\in C^\infty(V^*Q)$ which are analytic in fiber coordinates
on the vector bundle $V^*Q\to\bR$ at points
of its canonical zero section $q_j=p^j=0$.  

The algebra $\ol\cG(\bR)$ is provided with local Hermitian derivations
\be
&& \dr_t\f, \qquad \dr_k\f=\frac{i}{\hbar}[\dot\pi_k,\f], \qquad
\dr^k\f=-\frac{i}{\hbar}[\dot\vf^k,\f], \\
&& \dot\dr_k\f=\frac{i}{\hbar}[\pi_k,\f], \qquad
\dot\dr^k\f=-\frac{i}{\hbar}[\vf^k,\f], \qquad \f\in\ol\cG(\bR).
\ee
It is a desired prequantization algebra.

\bigskip 

\noindent 
{\bf III. PREQUANTIZATION OF JACOBI FIELDS}
\bigskip

In this Section, we aim to represent classical solutions of the 
Hamiltonian system and quantum Jacobi fields by Hermitian elements of the
prequantization algebra $\ol\cG(\bR)$. 

Let its Hamiltonian $\cH$ be a polynomial 
$\cH(t,q^k,p_k)$ of coordinates $(q^k,p_k)$ on $V^*Q$. This property is
coordinate-independent due to the linear transformations of these
coordinates. It should be emphasized that, being a part of the Hamiltonian
form (\ref{b4210}), the Hamiltonian
$\cH(t,q^k,p_k)$ is not a scalar under coordinate transformations, but
has the coordinate transformation law
\mar{j20}\beq
\cH'(t,q'^j,p'_j)=\cH(t,q^k,p_k) +p'_k(t,p_k)\dr_tq'^j(t,q^k).
\label{j20}
\eeq

Let $r$ be a local solution
of the classical Hamilton system, given by the local functions
(\ref{j40}).  Let they be analytic functions of
$a^j$, $b_j$ at values
$a^j=b_j=0$ and at $t$ from an open interval $U$ of $t=0$. We assign to this
solution the Hermitian elements 
\mar{j44}\beq
r^k(t,\vf^j, \pi_j),\qquad r_k(t,\vf^j,
\pi_j), \qquad r^k(0)=\vf^k, \qquad r_k(0)=\pi_k, \label{j44}
\eeq
of the algebra
$\ol\cG(U)$. They satisfy the equalities
\mar{j22}\beq
\dr_t r^k=(\dr^k\cH)(t,r), \qquad \dr_t r_k=-(\dr_k\cH)(t,r),
\label{j22}
\eeq
where $(\dr^k\cH)(t,r^j, r_j)$ and $(\dr_k\cH)(t,r^j,r_j)$ are also Hermitian
elements of the algebra $\ol\cG(U)$.

Thus, one can think of (\ref{j44}) as being the 
prequantization (\ref{j21}) of the classical solution (\ref{j40}) of a
Hamiltonian system. Note that classical solutions which differ from each other
in the initial values have the same prequantization (\ref{j44}). As will be
seen below, they correspond to different mean values of the operators
(\ref{j44}).
 
Turn now to prequantization of Jacobi fields.
Let us consider the equations 
\mar{j43}\bea
&& \dr_t\rho^k=\frac12[\rho^j(\dr_j\dr^k\cH)(t,r)+ 
(\dr_j\dr^k\cH)(t,r)\rho^{j*} + \label{j43a}\\
&& \qquad \rho_j(\dr^j\dr^k\cH)(t,r) +
(\dr^j\dr^k\cH)(t,r)\rho_j^*], \nonumber \\ 
&& \dr_t\rho_k=-\frac12[\rho^j(\dr_j\dr_k\cH)(t,r) + (\dr_j\dr_k\cH)(t,r)
\rho^{j*} + \label{j43b}\\
&& \qquad \rho_j(\dr^j\dr_k\cH)(t,r)
+(\dr^j\dr_k\cH)(t,r)\rho_j^*]. \nonumber
\eea
for elements $\rho^k$, $\rho_k$ of the prequantization algebra $\ol\cG(U)$.
They are similar to the equations (\ref{j42}) for classical Jacobi fields.
Let  
\mar{j45}\beq
\dot r^k(t,\vf^j, \pi_j, \dot\vf^j, \dot \pi_j),\qquad r_k(t,\vf^j,
\pi_j, \dot\vf^j, \dot \pi_j), \qquad \dot r^k(0)=\dot \vf^k, \qquad
\dot r_k(0)=\dot \pi_k,
\label{j45}
\eeq
be a local solution of the equations (\ref{j43a}) -- (\ref{j43b}) 
for $t\in V\subset U$. Point out the following two properties of such a
solution.

(i) Since the right-hand side of the equations (\ref{j43a}) -- (\ref{j43b}) is
Hermitian and the initial values are so, any solution (\ref{j45}) of these
equations is given by Hermitian elements of the prequantization algebra.

(ii) Using the equalities (\ref{j22}) and (\ref{j43a}) -- (\ref{j43b}), one can
easily justify that, if the elements  $r^k(t)$,
$r_k(t)$ (\ref{j44}) and the elements
$\dot r^k(t)$,
$\dot r_k(t)$ (\ref{j45}) of $\ol\cG(V)$, taken at some instant $t\in V$, obey
the commutation relations (\ref{j16a}) -- (\ref{j16c}), then the elements 
\be
&& r^k(t)+
\dr_tr^k(t)\Delta t, \qquad 
r_k(t)+\dr_tr_k(t)\Delta t, \\
&&  \dot r^k(t)+\dr_t\dot r^k(t)\Delta t, \qquad \dot
r_k(t)+ \dr_t\dot r_k(t)\Delta t, \qquad \Delta t\in \bR,
\ee
do so modulo terms
of order more than one in $\Delta t$. 
It follows that the elements (\ref{j44}) and (\ref{j45}) of the
prequantization algebra $\ol\cG(V)$  obey the commutation relations 
(\ref{j16a}) -- (\ref{j16c}) at any $t\in V$.

Note that, for many physical models, the prequantized
vertical-extended Hamiltonian
\mar{j47}\beq
\dr_V\cH(t,\vf^j,\pi_j,\dot\vf^j,\dot\pi_j)=
\dot\vf^j(\dr_j\cH)(t,\vf^i,\pi_i) +
\dot\pi_j(\dr^j\cH)(t,\vf^i,\pi_i) \label{j47}
\eeq
is Hermitian.  A glance at the transformation rule (\ref{j20})
shows that this property is coordinate-independent. In this case, a solution
(\ref{j45}) of the equations (\ref{j43a}) -- (\ref{j43b}) obey the matrix
equality
\be
\dr_t(\dot r^k,\dot r_k)= (\dot r^k,\dot r_k) M(t, \vf^j, \pi_j)
\ee
where
\be
&& M_j^k=(\dr_j\dr^k\cH)(t,r), \quad M^{jk}=(\dr^j\dr^k\cH)(t,r), \\
&& M_{jk}=(\dr_j\dr_k\cH)(t,r), \quad M^j_k=(\dr^j\dr_k\cH)(t,r).
\ee
Therefore, one can write such a solution as the time-ordered matrix exponent
\mar{j46}\beq
(\dot r^k,
\dot r_k)=(\dot\vf^k,\dot\pi_k)T\exp\left[\op\int_0^t  M(t', \vf^j, \pi_j)
dt'\right].
\label{j46}
\eeq
Moreover, it is readily observed that, when $r$ (\ref{j44}) is the
prequantization (\ref{j21}) of the classical solution (\ref{j40}), then $\dot
r$ (\ref{j46}) is exactly the prequantization (\ref{j18}) -- (\ref{j21}) of
the Jacobi field (\ref{j24}) if the latter obeys the required analyticity
condition.

Thus, the Hermitian elements (\ref{j44}), (\ref{j45}) provide prequantization
of solutions (\ref{j40}) of a classical Hamiltonian system and their Jacobi
fields (\ref{j41}). 

Note that, if the prequantized vertical-extended Hamiltonian (\ref{j47})
is Hermitian, 
$r$ (\ref{j44}) and $\dot r$ (\ref{j45}) are
solutions of the
evolution equations
\be
&& i\hbar \dr_t r^k=[r^k,\dr_V\cH(t,r,\dot r)], \qquad 
i\hbar \dr_t r_k=[r_k,\cH(t,r,\dot r)],\\
&& i\hbar \dr_t\dot r^k=[\dot r^k,\dr_V\cH(t,r,\dot r)],
\qquad  i\hbar \dr_t \dot r_k=[\dot r_k,\dr_V\cH(t,r,\dot r)]
\ee
with respect to the 
Hamiltonian $\dr_V\cH(t,r^j, r_j,\dot r_j,\dot r^j)$. Though written in a local
coordinate form, these equations are well behaved under coordinate
transformation due to the coordinate transformation law (\ref{j20}) of a
Hamiltonian. Indeed,  one can think of these equations as prequantization of
the covariant derivative of a section of the vertical momentum phase 
bundle 
$VV^*Q\to \bR$ with respect to the Hamiltonian connection
\be
&& \g_H=\dr_t+\dr^k\cH\dr_k -\dr_k\cH\dr^k +\dr_V\dr^k\cH\dot\dr_k -
\dr_V\dr_k\cH\dot\dr^k,\\
&&  \g_H\rfloor\bom_V= d H_V,
\ee 
on this fiber bundle.

\begin{rem}
Any first order dynamic equation 
\be
d_t q^k=\g^k, \qquad d_tp_k=\g_k
\ee
on the momentum phase bundle $V^*Q\to\bR$ can be seen as the Hamilton equations
(\ref{z740a}) for the Hamiltonian form
\be
H_V=\dot p_k(dq^k-\g^k) -\dot q^k(dp_k-\g_k)= 
\dot p_kdq^k -\dot q^kdp_k-(\dot p_k\g^k -\dot q^k\g_k)
\ee 
on the vertical momentum phase space $VV^*Q$.$^{4,12}$ One can apply the above
prequantization  construction to this Hamiltonian form, but the property (ii)
of a solution of the prequantization Jacobi equations (\ref{j43a}) --
(\ref{j43b}) need not take place.
\end{rem}

\bigskip 

\noindent 
{\bf IV. QUANTUM JACOBI FIELDS}
\bigskip

Now we aim to construct a representation of the elements $r(t)$ (\ref{j44}) and
$\dot r(t)$ (\ref{j45}) of the algebra $\ol\cG(\bR)$, describing
prequantization of classical solutions of a Hamiltonian system and their
Jacobi fields, by Hermitian operators in a Hilbert space. 
We use the fact that, in accordance with the property (ii) of a solution of
the prequantization Jacobi equations (\ref{j43a}) -- (\ref{j43b}), these
elements obey the commutation relations (\ref{j16a}) -- (\ref{j16b}) at any
instant
$t\in V\subset \bR$. 

Let $G^t$ denote the instant Lie algebra generated
by the elements 
\mar{j55}\beq
I, \quad r^k(t,\vf^k,\pi_k), \quad r_k(t,\vf^k,\pi_k), \quad \dot
r^k(t,\vf^k, \pi_k,\dot \vf^k, \dot \pi_k), \quad
\dot r_k(t,\vf^k, \pi_k,\dot \vf^k, \dot \pi_k) \label{j55}
\eeq 
at
$t\in V$. It has two Lie subalgebras $G^t_q$ and $G^t_p$ generated
respectively by the elements $\{r^k(t), \dot r_k(t),I\}$ and $\{r_k(t), \dot
r^k(t),I\}$. These subalgebras are isomorphic to an algebra of canonical
commutation relations (a CCR-algebra), but not in a canonical way. 
These isomorphisms
are defined by a fibre metric $g$ on the vector bundle $Q\to\bR$
 as follows. Given a trivialization $Q|_V=V\times
\bR^m$ associated with local coordinates $(t,q^k)$, let $\dl_{kj}$ be the
Euclidean metric on $\bR^m$.  Put 
\be
r_k^g(t) =\frac1\hbar g_{kj}\ r^j(t)=\frac1\hbar r^k(t), \qquad \dot
r_k^g(t)=\frac1\hbar g_{kj}\dot r^j(t)=\frac1\hbar \dot r^j(t). 
\ee
Then $\{r_k^g(t), \dot r^k(t),I\}$ and $\{r^k(t), \dot r_k^g(t),I\}$ obey
respectively the commutation relations 
\be
[r^g_k(t),r^g_j(t)]=[\dot r_k(t),\dot r_j(t)] =0, \qquad
[\dot r_k(t), r^g_j(t),] =-i \dl_{kj} I
\ee
and 
\be
[\dot r^g_k(t),\dot r^g_j(t)]=[r_k(t), r_j(t)] =0, \qquad
[r_k(t),\dot r_j^g(t)] =-i \dl_{kj} I
\ee
which are exactly the commutation relations of the standard
Heisenberg--Weyl CCR-algebra modelled over the finite-dimensional 
Euclidean space
$\bR^m$. Therefore, one can obtain representations of the instant algebras 
$G^t_q$ and $G^t_p$ from the well-known representations of this CCR-algebra.
Of course, these representations depend on the choice of a coordinate chart on
the configuration bundle $Q\to\bR$, but they are equivalent by virtue of the
well-known Stone-von Neumann uniqueness theorem.

There are different variants of a representation of the Heisenberg--Weyl
CCR-algebra (see, e.g., Ref. [23]). We choose its Shr\"odinger representation
where the operators
$r^k(t)$ of the instant algebra $G^t_q$ and the operators $r_k(t)$ of the
instant algebra
$G^t_p$ have a complete set of eigenvectors.$^{21}$ The representation of the
Lie algebra
$G^t$ is obtained as the (topological) tensor product of representations 
of the Lie algebras $G^t_q$ and $G^t_p$. Moreover, it suffices to construct
a representation of the instant algebra $G^{t=0}=\cG_0$ with generators
$\{\vf^k,\pi_k,\dot\vf^k,\dot\pi_k,I\}$ which fulfill the
commutation relations (\ref{j16a}) -- (\ref{j16c}). Then one can obtain a
representation of the instant algebra $G^t$, $t\in V$ as a subalgebra of
the enveloping algebra $\ol\cG_0$ due to the splitting 
\be
Q|_V=V\times Q_{t=0}=V\times\bR^m.
\ee

A desired representation of the Lie algebra $\cG_0$ is defined in the Hilbert
space $L^2(\bR^{2m}, \mu_g)$ of complex functions $f(x^k,y_k)$ on
$\bR^{2m}=Q^*_0\oplus Q_0$ which are square-integrable with respect to the
Gaussian measure
\be
d\m_g=\pi^{-m}\exp\left[-\op\sum_k (x^k)^2 - \op\sum_k (y_k)^2\right]
d^mx d^my.
\ee    
This representation is given
by the operators
\mar{j51}\beq
\vf^k=i\hbar\frac{\dr}{\dr y_k}, \qquad \pi_k=-i\hbar \frac{\dr}{\dr x^k},
\qquad \dot \vf^k= x^k, \qquad \dot \pi_k= y_k \label{j51}
\eeq
on the dense subspace of smooth functions of $L^2(\bR^{2m}, \mu_g)$.
It is readily observed that they are Hermitian operators with respect to the
Hermitian form
\be
\lng f|f'\rng=\int f\ol f' d\m_g
\ee
on $L^2(\bR^{2m}, \mu_g)$. In this representation, the operators $\vf^k$ and
$\pi_j$ have the common eigenstates
\mar{j53}\beq
f_{a,b}=\exp\left[\frac{i}{\hbar}(b_kx^k - a^k y_k)\right]. \label{j53}
\eeq 

Given the representation (\ref{j51}) of the algebra $G^0$, the representation
of the instant algebra $G^t$ with the generators (\ref{j55})
is given in the same Hilbert
space
$L^2(\bR^{2m}, \mu_g)$ by the operators
\mar{j56}\bea
&& \br^k(t)= r^k(t,i\hbar\frac{\dr}{\dr y_k},-i\hbar \frac{\dr}{\dr x^k}),
\qquad \br_k(t)=r_k(t,i\hbar\frac{\dr}{\dr y_k},-i\hbar \frac{\dr}{\dr x^k}),
\label{j56a}\\ 
&& \dot \br^k(t)=\dot
r^k(t,i\hbar\frac{\dr}{\dr y_k}, -i\hbar \frac{\dr}{\dr x^k},x^k, y_k),
\qquad
\dot\br_k(t)=\dot r_k(t,i\hbar\frac{\dr}{\dr y_k}, -i\hbar \frac{\dr}{\dr
x^k},x^k, y_k). \label{j56b}
\eea 
The operators (\ref{j56a}) also have eigenstates $f_{a,b}$
(\ref{j53}). Their eigenvalues at the eigenstate $f_{a,b}$ are exactly the
values at the instant $t$ of the classical solution (\ref{j40}) with the
initial values
$a^k$ and $b_k$.

One can think of the operators (\ref{j56b}) as being the operators of quantum
Jacobi fields at the instant
$t$. Indeed, the operators 
\mar{j57}\beq
R(t,\bt_k)=\exp\{\frac{i}{\hbar}\bt_k\dot \br^k(t)\}, \qquad
R(t,\al^k)=\exp\{ -
\al^k\frac{i}{\hbar}\dot
\br_k(t)\} \label{j57}
\eeq
of the Weyl Lie group of the Lie algebra $G_t$
obey the commutation relations
\be
&& [\br_j(t), R(t,\bt_k)]= \bt_j R(t,\bt_k), \qquad [\br_j(t), R(t,\al^k)]=0,\\
&& [\br^j(t), R(t,\al^k)]= \al^j R(t,\al^k), \qquad [\br^j(t), R(t,\bt_k)]=0.
\ee
It means that, given an eigenstate $f_{a,b}$ of the operators 
$\br^j(t)$ and $\br_j(t)$ with the eigenvalues $r^j(t,a,b)$ and $r_j(t,a,b)$,
the vectors
$R(t,\bt_k)f_{a,b}$, 
$R(t,\al^k)f_{a,b}$ are also eigenvectors of these operators with the
eigenvalues $r^j(t,a,b)$, $r^j(t,a,b)+\al^j$ for $\br^j(t)$ and 
$r_j(t,a,b)+\bt_j$, $r_j(t,a,b)$ for $\br_j(t)$.
In particular, the operators $R(0,\bt_k)$ and $R(0,\al^k)$
 send the eigenstate $f_{a,b}$ to the
eigenstates
$f_{a,b+\bt}$ and $f_{a+\al,b}$, respectively. 

Thus, one can say
that the operators (\ref{j57}) of quantum Jacobi fields  perform a transition
 between the classical solution of the original Hamiltonian system.

Now we can easily construct the representation of elements $r(t)$ (\ref{j44}) 
and $\dot r(t)$ (\ref{j45}) of the prequantization algebra $\ol\cG(\bR)$
restricted to the interval $V=(-\ve,\ve)$ of $\bR$. This representation is
defined  in the Hilbert space $L^2((-\ve,\ve)\times\bR^{2m},
\mu)$ of complex functions $f(t,x,y)$
on  $(-\ve,\ve)\times \bR^{2m}$ which are square-integrable
with respect to the measure $d\m=(2\ve)^{-1}d\m_gdt$. The Hermitian form on
this Hilbert space reads
\be
\lng f|f'\rng=\int f\ol f' d\m.
\ee
The representation operators take the form (\ref{j56a}) -- (\ref{j56b}) where
$t$ now is a variable. In particular, the functions $f_{a,b}$ (\ref{j53}) are
vectors of this representation, but not eigenvectors of the operators
$\br(t)$. 

\begin{rem}
If the prequantization operators of classical solutions and Jacobi fields
are defined on $\bR$, one can construct their representation in the Hilbert
space $L^2(\bR^{2m+1},
\mu=\m_t\times\m_g)$ by a choice of some measure $\m_t$ of total mass 1 on
$\bR$.
\end{rem}

\bigskip 

\noindent 
{\bf V. QUANTUM JACOBI FIELDS OF THE HARMONIC OSCILLATOR}
\bigskip

In order to illustrate the above construction, let us provide quantization of
Jacobi fields of a one-dimensional harmonic oscillator. 

Its configuration
space $Q$ is the line bundle $\bR^2\to\bR$ coordinated by $(t,q)$. The
corresponding momentum phase space $V^*Q$ is the plane bundle $\bR^3\to\bR$
coordinated by $(t,q,p)$, while the vertical momentum phase bundle is
$\bR^5\to\bR$ endowed with holonomic coordinates $(t,q,p,\dot q,\dot p)$.
The Hamiltonian form $H$ of a harmonic oscillator reads
\mar{j70}\beq
H=pdq-\cH dt, \qquad \cH=\frac12(p^2+ \om^2q^2), \label{j70}
\eeq
where we put the oscillator mass equal 1 for brevity. 
The vertical extension (\ref{j5}) of this Hamiltonian form is
\be
H_V=pd\dot q +\dot p dq -\cH_Vdt, \qquad \cH_V=\dot p p + \om^2\dot qq.
\ee
It leads to the Hamilton equations
\mar{j61}\bea
&& d_tq= p, \qquad d_t p= -\om^2q, \label{j61a}\\
&& d_t\dot q= \dot p, \qquad d_t \dot p=-\om^2\dot q. \label{j61b}
\eea
The Hamilton equations (\ref{j61a}) have a familiar solution
\mar{j62}\beq
q(t)=q_0\cos\om t + p_0\om^{-1}\sin \om t, \qquad p(t)=-q_0\om\sin\om t +
p_0\cos\om t, \label{j62}
\eeq
where $q_0$ and $p_0$ are the initial values. The Hamilton equations
(\ref{j61b}) for Jacobi fields have the
similar solution
\mar{j63}\beq
\dot q(t)=\dot q_0\cos\om t + \dot p_0\om^{-1}\sin \om t, \qquad
\dot p(t)=-\dot q_0\om\sin\om t + \dot p_0\cos\om t. \label{j63}
\eeq

The prequantization algebra $\ol\cG(\bR)$  over the ring
$\bC(\bR)$ of smooth complex functions on $\bR$ is generated by the elements
$\{\vf,\pi,\dot\vf,\dot \pi, I\}$ which obey the commutation relations
\mar{j65}\beq
 [\vf,\pi]=[\dot \vf,\dot\pi]=[\vf,\dot\vf]=[\pi,\dot \pi]=0, \qquad
 [\pi,\dot\vf]=[\dot \pi,\vf]=-i\hbar I. \label{j65}
\eeq
The prequantization of the solution (\ref{j62}) of a classical harmonic
oscillator and the Jacobi fields (\ref{j63}) by Hermitian elements of the
algebra
$\ol\cG(\bR)$ reads
\mar{j64}\bea
&& q(t)=\vf\cos\om t + \pi\om^{-1}\sin \om t, \qquad p(t)=-\vf\om\sin\om t +
\pi\cos\om t, \label{j64a}\\
&& \dot q(t)=\dot \vf\cos\om t + \dot \pi\om^{-1}\sin \om t, \qquad
\dot p(t)=-\dot \vf\om\sin\om t + \dot \pi\cos\om t. \label{j64b}
\eea
It is readily observed that they obey the commutation relations
\be
&& [q(t),p(t)]=[\dot q(t),\dot p(t)]=[q(t),\dot q(t)]=[p(t),\dot p(t)]=0, \\
&& [p(t),\dot q(t)]=[\dot p(t),q(t)]=-i\hbar I.
\ee
Note that the prequantized vertical-extended Hamiltonian
\be
\cH_V=\dot \pi \pi + \om^2\dot \vf\vf 
\ee
is Hermitian.

Following the general scheme, we construct the representation of the Lie
algebra (\ref{j65}) in the Hilbert space $L^2(\bR^2,\m_g)$ of 
complex functions $f(x,y)$ on
$\bR^2$ which are square-integrable with respect to the
Gaussian measure
\be
d\m_g=\pi^{-1}\exp[-x^2 - y^2] dx dy.
\ee    
This representation is given
by the operators
\be
\vf=i\hbar\dr_y, \qquad \pi=-i\hbar\dr_x,
\qquad \dot \vf= x, \qquad \dot \pi= y. 
\ee
Accordingly, the elements (\ref{j64a}) -- (\ref{j64b}) are represented by the
operators
\mar{j68}\bea
&& \bq(t)=i\hbar\dr_y\cos\om t  - i\hbar\dr_x\om^{-1}\sin \om t, \qquad
\bp(t)=-i\hbar\dr_y \om\sin\om t -
i\hbar\dr_x \cos\om t, \label{j68a}\\
&& \dot \bq(t)=x\cos\om t + y\om^{-1}\sin \om t, \qquad
\dot \bp(t)=-x\om\sin\om t + y\cos\om t. \label{j68b}
\eea
Considered at a given instant $t\in\bR$, these operators act in the Hilbert
space $L^2(\bR^2,\m_g)$, and the operators  (\ref{j68a}) have the eigenstates
\be
f_{q_0,p_0}=\exp\left[\frac{i}{\hbar}(p_0x - q_0 y)\right]. 
\ee 
Their eigenvalues at the eigenstate $f_{q_0,p_0}$ are exactly the values of
the classical solution (\ref{j62}) at the instant $t$.

Treated as operator functions of $t\in\bR$, the operators (\ref{j68a}) --
(\ref{j68b}) act in the Hilbert space $L^2(\bR^3,\m)$ of 
complex functions $f(t,x,y)$ on
$\bR^3$ which are square-integrable with respect to the
Gaussian measure
\be
d\m=\pi^{-3/2}\exp[-t^2-x^2 - y^2] dt dx dy.
\ee    

It is readily observed that the Hamiltonian $\cH_2$ (\ref{j5'}) for linear
deviations of a harmonic oscillator coincides with the quadratic Hamiltonian
(\ref{j70}) of this oscillator after the replacement of $q$, $p$ with $Q$,
$P$. The Hamiltonians $\cH_1$ and $\cH_2$ (\ref{j5'}) for a harmonic
oscillator are independent.  Therefore, the quantization of Jacobi fields of
a harmonic oscillator and that of linear deviations of a
harmonic oscillator (i.e., the standard quantization of a harmonic oscillator
itself) are independent. It follows that quantizations of a harmonic
oscillator over different classical solutions coincide with each other, and
quantum Jacobi fields do not transform the states of a quantum Harmonic
oscillator.

\end{document}